\documentclass[structabstract]{aa}  
\usepackage{graphicx}
\usepackage{txfonts}
\usepackage{psfig}
\begin{document}
   \title{Diffractive and refractive timescales at 4.8 GHz in PSR B0329+54} 


   \author{W. Lewandowski\inst{1,3},
          J. Kijak
          \inst{1},
          Y. Gupta
          \inst{2},
          \and
          K. Krzeszowski
          \inst{1},
          }

   \offprints{W. Lewandowski}

   \institute{Kepler Institute of Astronomy, University of Zielona G\'ora,
              Lubuska 2, Zielona G\'ora, PL-65-265 Poland \\
              \email{boe@astro.ia.uz.zgora.pl}
         \and
            National Centre for Radio Astrophysics, TIFR, Pune
              University Campus, Pune, India 
          \and
            Toru\'n Centre for Astronomy of the Nicolaus Copernicus University,
              Department of Radio Astronomy, Gagarina 11, Toru\'n,
                     PL-87-100 Poland 
}

   \date{Received ....; accepted ....}

 
  \abstract
 {}
{We present the results of flux density monitoring of PSR~B0329+54 at 
the frequency of 4.8~GHz using the 32-meter TCfA radiotelescope.
The observations were conducted between 2002 and 2005. The main goal of the
project was to find interstellar scintillation (ISS) parameters for the pulsar at the frequency
at which it was never studied in detail. To achieve this the 20 observing 
sessions consisted of 3-minute integrations which on average lasted 24 hours. This gave us sufficient 
sensitivity to all types of flux density variations over a wide range of timescales. 
The character of the observations makes our project unique amongst other ISS oriented
observing programs, at least at high frequency.}
{Flux density time series obtained for each session were analysed using structure functions. 
For some of the individual sessions as well as
for the general average structure function we were able to identify two distinctive timescales present,
the timescales of diffractive and refractive scintillations. 
To the best of our knowledge, this is the first case when both scintillation timescales, 
$t_{\rm DISS} = 42.7$ minutes and  $t_{\rm RISS} = 305$ minutes, were observed simultaneously in
a uniform data set and estimated using the same method.}
{The obtained values of the ISS parameters combined with the data
found in the literature allowed us to study the frequency dependence 
of these parameters over a wide range of observing frequencies, which is crucial for understanding 
the ISM turbulence. We found that the Kolmogorov spectrum is not best suited for describing 
the density fluctuations of the ISM, and a power-law
spectrum with $\beta =4$ seems to fit better with our results. We were 
also able to estimate the transition frequency (transition from strong to weak scintillation regimes) 
as $\nu_c = 10.1$~GHz, much higher than was previously predicted. We were also able to estimate the 
strength of scattering parameter $u=2.67$ and the Fresnel scale as $r_F = 6.7 \times 10^8$ meters.} 
{}


\authorrunning{W. Lewandowski~et~al.}
\titlerunning{Two scintillation timescales for 0329+54}

   \maketitle


\section{Introduction}

Scintillations of the pulsar signal is a phenomenon that was identified shortly 
after the discovery of pulsars themselves (Hewish~et~al. \cite{hewish}). From an 
observer point of view pulsars are point-like sources, and presumably are intrinsically
stable sources of radiation. Their radiation, however, undergoes scattering process in the 
interstellar plasma on its way to the observer. 

Depending on the strength of the scattering that the radiation of a given pulsar undergoes, 
the interstellar scintillations (ISS) will result in different 
phenomena observed for the pulsar (see Lorimer \& Kramer, \cite{lori}, and the references therein for a recent review).
For a nearby pulsar, or when the radiation encounters only small amounts of the scattering
material (and if the observing frequency is higher than ca. 1~GHz), the disturbances of the wavefront 
will be minimal, resulting in {\it weak scintillations}.

In case of strong scattering (a distant pulsar and/or low observing frequency), the phase perturbations
become large, which leads to a strong modulation of the pulsar signal. In this {\it strong scintillation regime} 
one has to  take both diffraction and refraction effects into account. The {\it diffractive interstellar scintillations}
arise from small ($10^6 \div 10^8$ meter) inhomogeneities in the interstellar medium, which introduce 
observed flux density variations with the timescale that can be estimated as $t_{\rm DISS} \sim f^{1.2} \  d^{0.6}$
(assuming a Kolmogorov spectrum of the ISM turbulence, see Rickett \cite{rick}; Romani~et~al. \cite{romani}).

The {\it refractive interstellar scintillations} (RISS; Sieber \cite{sieb}; Rickett at al. \cite{rick84})
arise from larger scale density fluctuations ($10^{10} \div 10^{13}$ meters). As theories predict, the 
RISS timescale decreases with observing frequency, roughly as $t_{\rm RISS} \sim f^{-2.2} \ d^{1.6}$.

For any given pulsar the values of the mentioned parameters will hence change with the observational frequency.
Starting from very low frequencies, we will have short but increasing DISS timescales, and very long but decreasing RISS 
timescales, which will be accompanied by increasing RISS modulation (DISS modulation will be always
close to unity). The rate at which these parameters change can be used to decide between various theories describing 
the ISM turbulence spectrum, because their frequency dependence power-law predictions are tied to the model 
used (see Romani~et~al. \cite{romani}, Bhat~et~al. \cite{bhat}).

As the observing frequency increases, both scintillation timescales should be closer, and the refractive 
modulation index should increase towards unity (which is best illustrated by Lorimer \& Kramer, \cite{lori}, on their 
Figure~4.5b). At a certain frequency, called the {\it transition frequency} both timescales become equal, refractive
scintillations reach maximum modulation and the character of the scintillations in general changes. This is 
the frequency at which the transition from the strong to weak scintillation regimes happens.

Most of the pulsar scintillation observations conducted so far were performed at the frequencies of 1.5~GHz and below, 
there is only a handful of papers describing ISS observations at higher frequencies 
(Malofeev~et~al. \cite{malo}; Kramer~et~al. \cite{kram}; Shishov et~al. \cite{shis}). The attempts 
to estimate the spectrum of the 
ISM turbulence via the measurement of ISS parameters were limited to the frequency range from ca. 
$\sim 100$~MHz to 1.5~GHz. This is both because pulsars are steep spectrum sources and are considerably 
weaker at higher frequencies, and the increasing DISS timescale which means one needs longer 
data sets to be able to perform measurements in a way similar to what is done at low frequencies. 
Both of these reasons result in a much longer 
observing time to be able to obtain reliable measurements. Because there are only 
few radiotelesopes capable of high-frequency observations, projects like this are very hard to implement.

The way to at least partially overcome these obstacles is to limit the observations to very strong pulsars, 
which in turn lowers the telescope-size requirements. One of the best candidates for these observations is definitely 
PSR~B0329+54, the strongest radio pulsar in the northern hemisphere. Owing to its brightness, it was observed
by many various ISS projects in the past (see the {\it Discussion} section in this paper, and Table~\ref{tab3} 
for a full list of references). Amongst them was a project conducted by Malofeev~et~al.~(\cite{malo}), which involved 
fairly short (ca. 130 minutes) observations at 4.75 and 10.55~GHz, based on which the authors concluded 
that for PSR~B0329+54 (with DM = 26.833) the transition frequency is ca. 3~GHz. This would mean that at higher 
frequencies this pulsar should be in the weak scintillation regime, and this statement was contradicting our 
personal experience with that source. The flux density of PSR~B0329+54 measured for the purposes of estimating 
pulsar spectrum (see Maron~et~al. \cite{maro00} for an overview of that project) was showing very strong 
variations, and in individual measurements it ranged from a few to over a hundred milli-Janskys.

To solve this problem we decided to perform a long-time monitoring project of PSR~B0329+54 at the frequency
of 4.8~GHz. We were kindly given a chance of extensive use of Toru\'n Centre for Astronomy 32-meter radiotelescope
(TCfA, located near Toru\'n, Poland), which is equipped with a cooled 5~GHz receiver and the Penn State Pulsar Machine~II. 
We used it to perform numerous long-duration observing sessions between 2002 and 2005. This paper summarizes 
our results.


\section{Observations and analysis}

\begin{figure*}

\ \put(0,170)
{\includegraphics{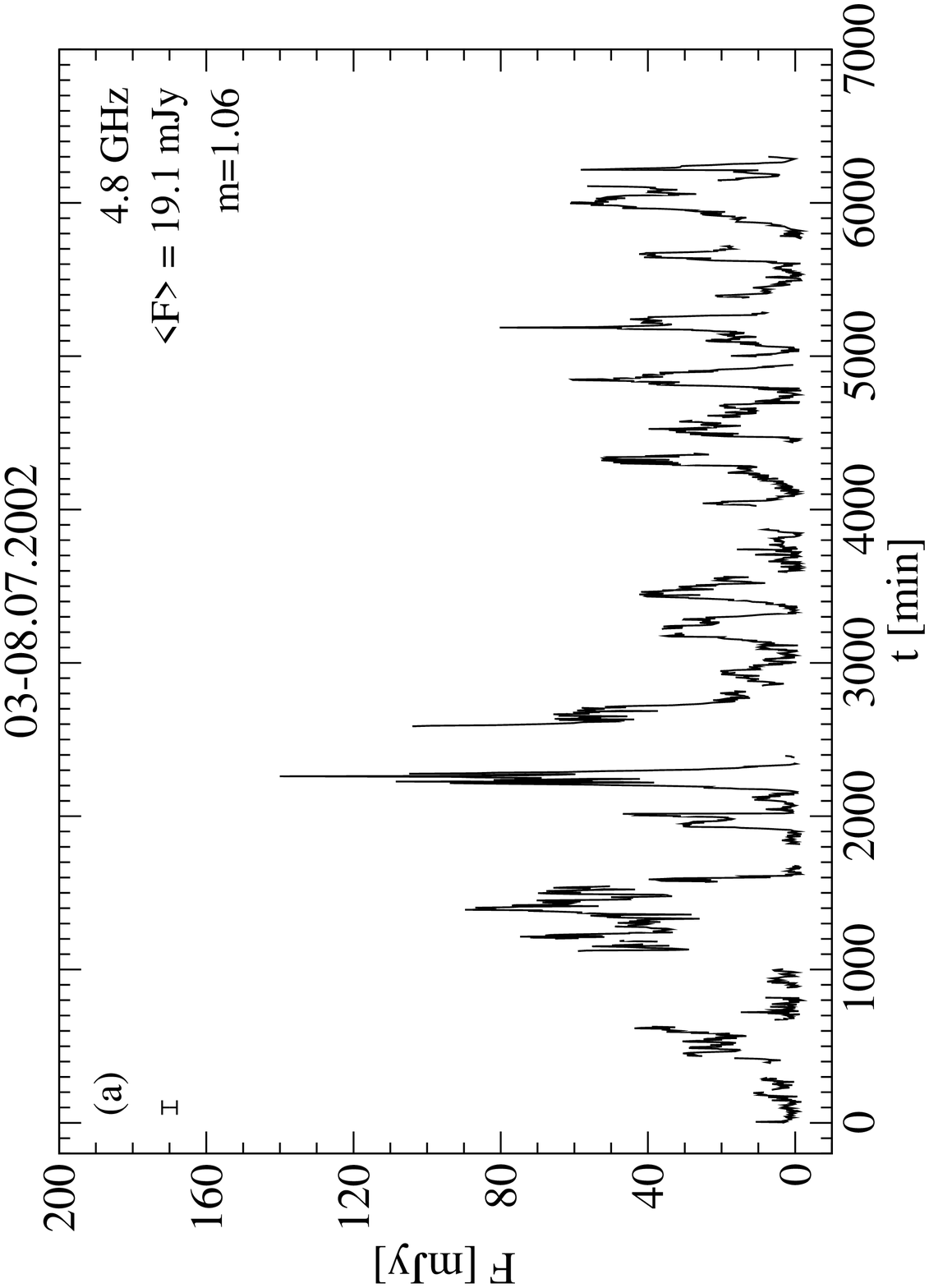}}
\ \put(220,170)
{\includegraphics{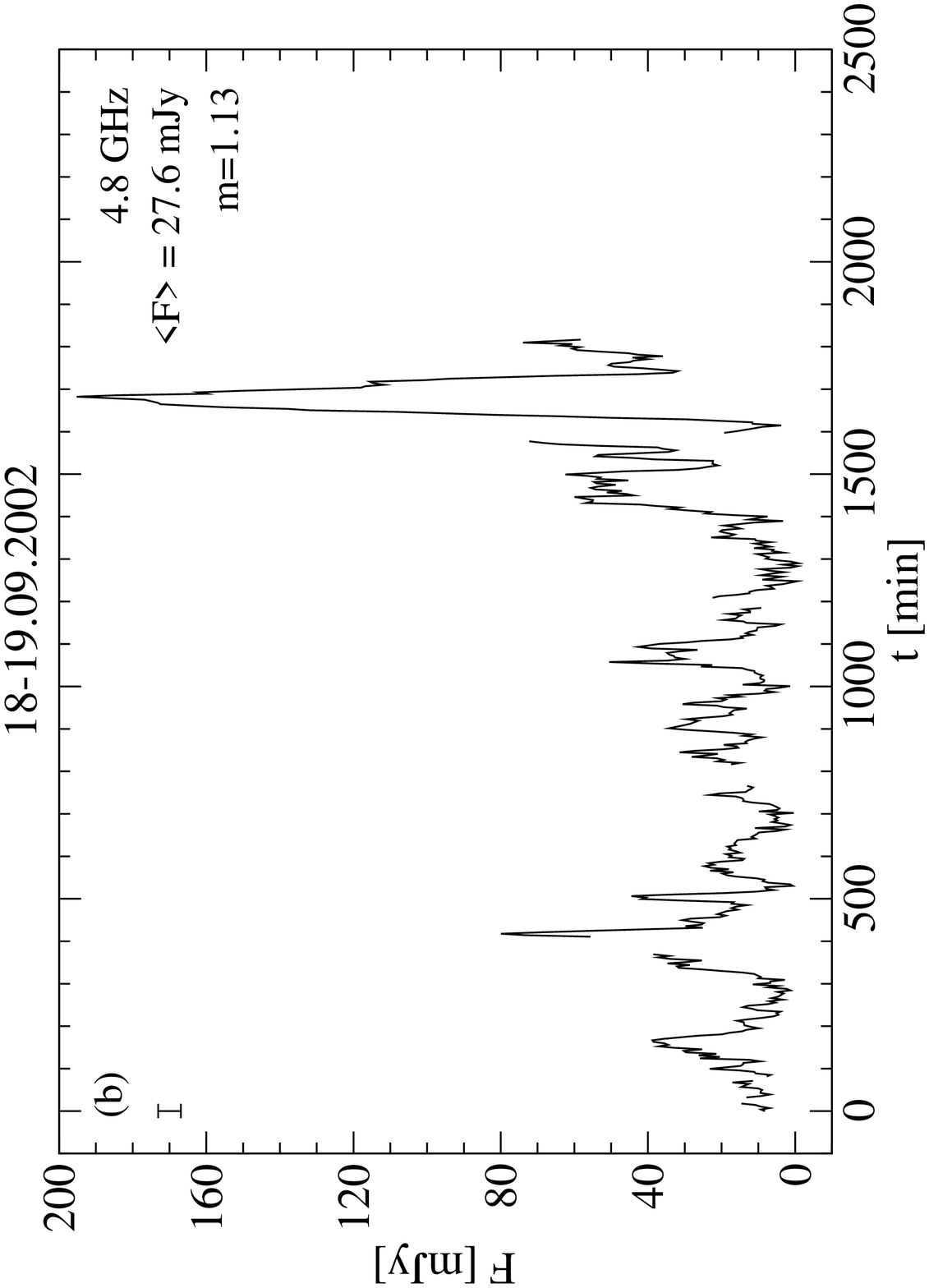}}
\ \put(0,0)
{\includegraphics{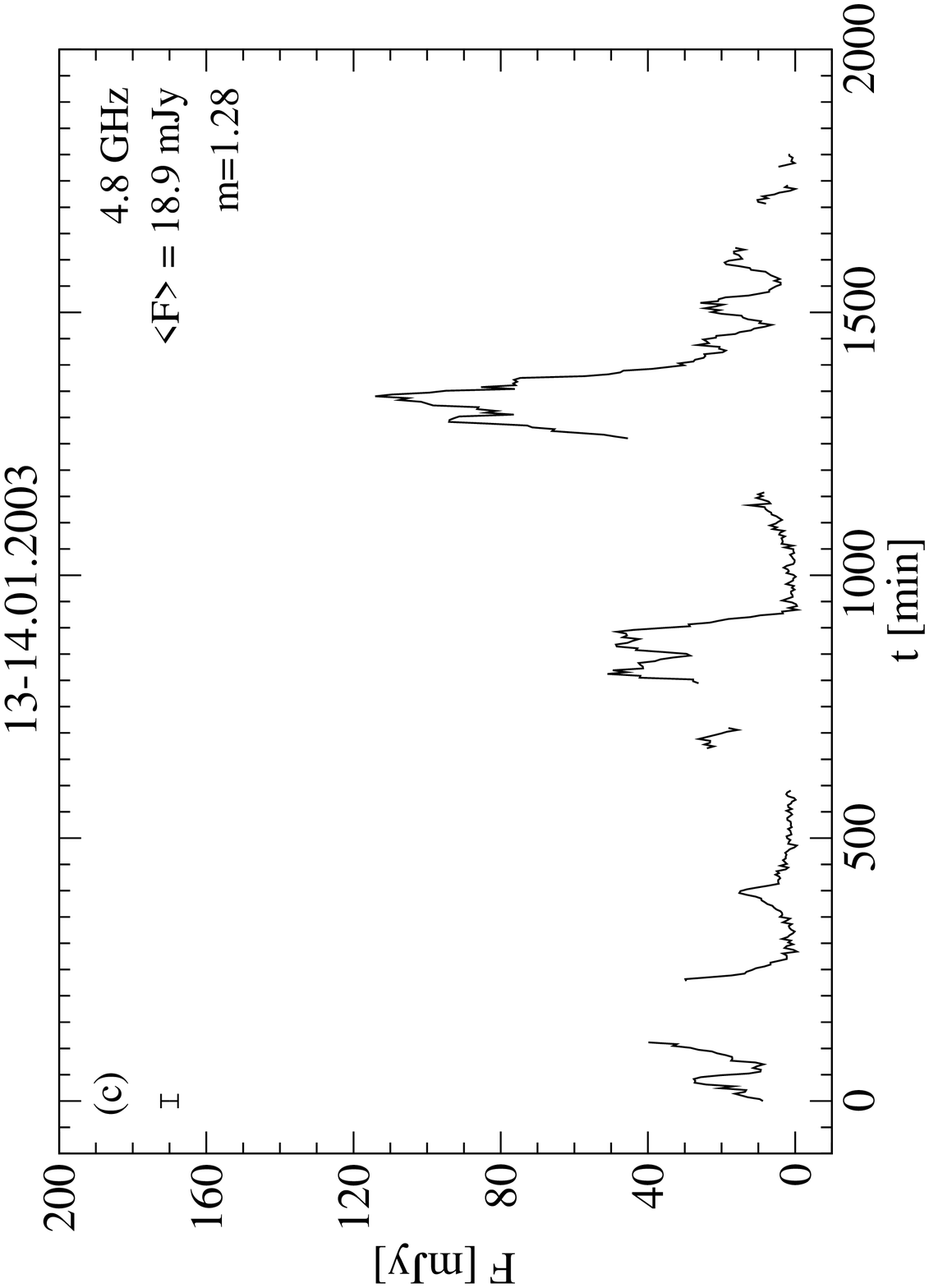}}
\ \put(220,0)
{\includegraphics{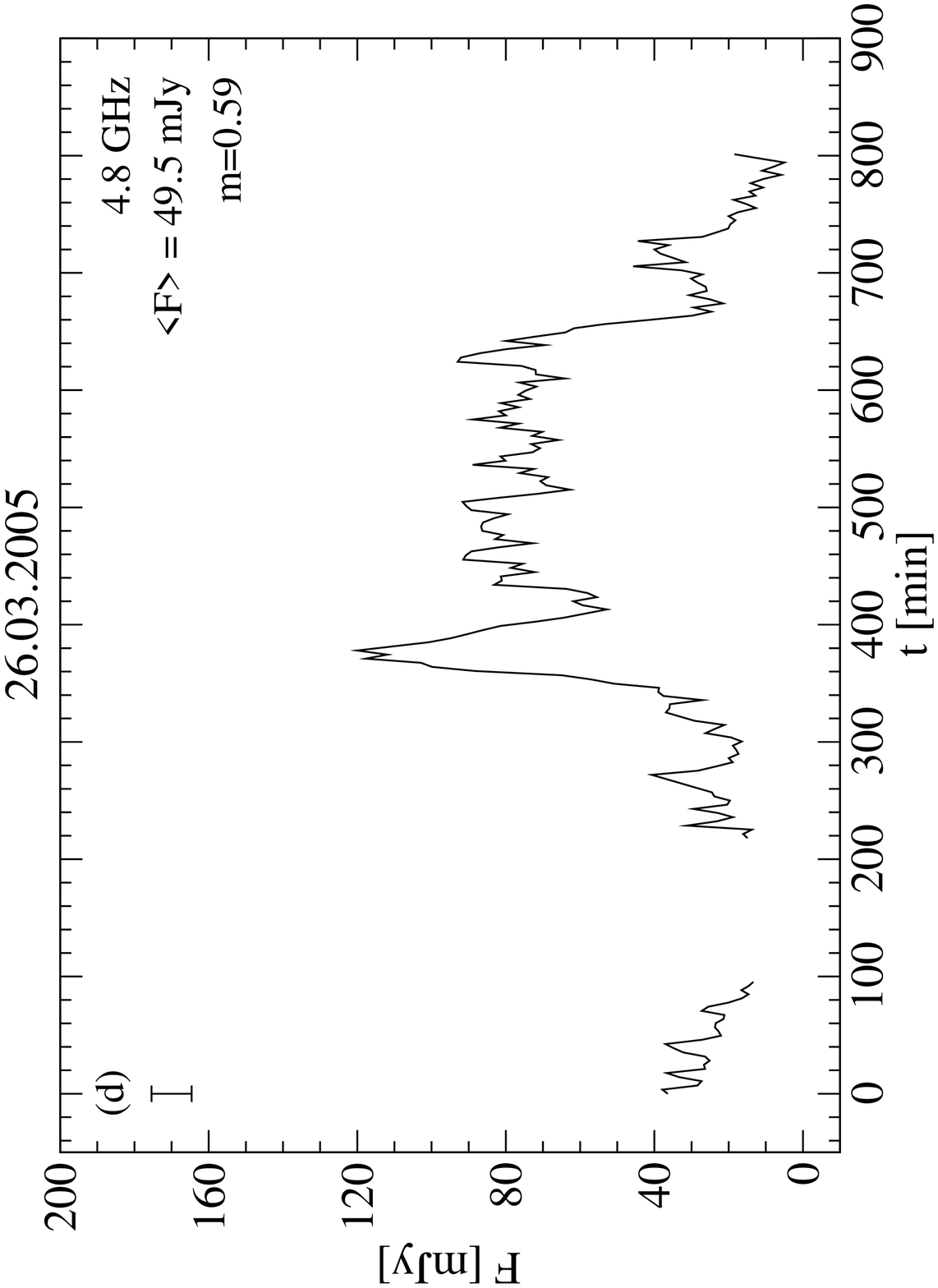}}
\vskip70mm
      \caption{Top left plot: flux time series for the longest observation time (5
              days).  Top right: flux time series with the largest flux variations.
              Bottom left: one of the unusual flux time series (gaps and some 
              uncharacteristic flux behaviour). Bottom right: strange flux variations
              characteristic for the last two of our sessions. The bar in the top-left corner of 
              each subplot represents $1 \sigma$ flux error for each session.
              \label{fig1}}
   \end{figure*}

We have carried out a flux density monitoring of PSR B0329+54 during three
years using the 32-meter Toru\'n Centre for Astronomy radiotelescope.
Observations were performed between July 2002 and June 2005 at the frequency
of 4.8 GHz. Twenty epochs 
of flux density monitoring at 4.8 GHz have been collected (see Table~1). Typical 
observations lasted about 12 - 30 hours in 3 minute scans. During the first two years 
of our campaign each epoch was at least 24 hours; for the third year we observed this pulsar
only occasionally, for about 12 hours each epoch. In addition, the first observing session 
lasted a full 5 days, and some other were almost 2 day long.

The total intensity signal was obtained using the Penn State Pulsar Machine~II (PSPM~II).
This pulsar back-end is basically a 64 channel fast radio-spectrometer, which was 
designed to conduct timing observations and search for pulsars 
(see Konacki~et~al. \cite{kon99}). The PSPM~II is currently equipped with 
additional system that allows mean flux calibration.
A dual-circular polarization signal from the total bandwidth of 192~MHz is fed into 
the backend, and then split into 64 channels (of 3~MHz width each, separately
for each polarization). Signals from both polarizations in each channel are then added, 
and sampled with a rate of 4096 samples per period, thanks to a system that allows
dynamical changes of the sampling rate during the observation. This makes data much 
easier to integrate over time, which in fact is done in real-time. 
At the same time the noise diode signal is injected once per pulsar period, always 
on the same phase, which was at least 0.3 different from the actual pulse phase.
As a result we have  got 64 integrated pulse profiles with a calibration mark  
for each single observation. Data from all channels were dedispersed and 
integrated over the whole bandwidth in an off-line procedure.

To obtain flux densities, we carried out regular calibration 
measurements using the same signal of a noise diode, which was compared to the
flux density of known continuum calibrators. Pulsar observations were
interrupted regularly for the calibration procedure roughly every 6-8 hours (such calibration
breaks took about one to one and a half hours), and sometimes there were unforeseen 
breaks owing to the telescope failures and hardware
problems.


\subsection{Flux density measurements \label{flux}}

Figure~\ref{fig1} shows four of the total 20 flux time-series as observed during our campaign. 
Subplot~{\bf (a)} represents the flux variation during the longest observing session, i.e. 
the first epoch in our campaign, that lasted for five days. Evidently, flux variations 
happen at more than one timescale; the detailed analysis is presented in the next sections. 
Subplot~{\bf (b)} shows the epoch with the highest measured flux value in our campaign. 
The maximum averaged flux for a single 3-minute integration reaches almost 200~mJy, which means 
more than 10-times the average value ($\left<F\right>=17.4$~mJy, see Table.\ref{t1}). This value is also close 
to the average flux of PSR~B0329+54 at 1400~MHz (203~mJy according to ATNF 
pulsar catalogue). An example of one of the worst quality sessions is shown in subplot~{\bf (c)}.
There are many gaps in the data taken at that epoch, some are the regular calibration breaks, 
but others are caused bye the telescope/hardware failures. With all those gaps and some 
uncharacteristic flux variation it was not possible to conduct a proper analysis for this epoch 
(confront Table.\ref{t1} and Section.\ref{sect_sf} for detailed explanation). The last subplot {\bf (d)} shows
the flux variations for the penultimate epoch. Again, the flux variations are not typical for 
our data, the value stays well above the average for a very long time (more than 300 minutes). 
This kind of variability makes some kinds of statistical analyses unreliable, i.e. we were not 
able to find the scintillation timescales because the structure function of the flux time series
for this data set does not saturate. At the same time one has to note that this behaviour is quite 
typical for PSR~B0329+54, at least in our data. If one confronts the flux variations from 
subplot~{\bf (d)} with the longest time series (subplot~{\bf (a)}), one can easily find a matching 
behaviour - clearly visible between 1000 and 1700 minutes during the 5-day session. For the epoch 
shown in subplot~{\bf (d)}, which lasted only for 800 minutes, we were just ``unlucky'' to catch 
this kind of variation and not the regular scintillation pattern. The same applies for the last 
session, with average flux value also exceeding the general average calculated from all sessions
(see Table.\ref{t1}).

Because these last two sessions were so untypical for this pulsar, we decided to calculate the general average 
flux value, and also the average modulation index, with and without them. Both results are shown in
Table.\ref{t1}. This table presents the results of the mean flux measurements, averaged over
each observing session $\left<F\right>$, along with their standard deviations ($\bar \sigma_{<F>}$),
which also include the contribution from the calibration source uncertainty.
We also present the flux modulation indices calculated for individual sessions (see
Equation.\ref{mod1}), these will be discussed in detail in the following sections of this paper.

\begin{table}[t]
\caption{Average flux density measurements for each epoch and modulation index
  (as calculated from Eq.\ref{mod1}). \label{t1}}
\begin{tabular}{lrrcc}
\hline\hline
Date~~~~~~~  & $\rm t_{~obs}$~& $<F>$ &  $\bar \sigma_{<F>}$ &  $m$  \\
             &{\scriptsize  (min)} & {\scriptsize (mJy)} &{\scriptsize (mJy)} & \\
\hline
2002-Jul-03-08$^a$  &   6300.40 &   19.1  &    2.0  &   1.06  \\
2002-Aug-22-23$^a$  &   1454.30 &   17.4  &    2.1  &   1.32  \\
2002-Sep-18-19$^a$  &   1816.80 &   27.6  &    3.1  &   1.13  \\
2002-Nov-19-20      &   1731.50 &   14.0  &    1.5  &   0.71  \\
2003-Jan-13-14$^b$  &   1801.30 &   18.9  &    2.3  &   1.28  \\
2003-Feb-06-07      &   1277.00 &   20.6  &    2.3  &   0.83  \\
2003-Feb-19-21      &   2650.30 &   11.9  &    1.3  &   0.77   \\
2003-Mar-16-18      &   1707.50 &    6.6  &    0.7  &   0.88  \\
2003-May-01-02$^b$  &   1852.30 &   11.5  &    1.2  &   0.91  \\
2003-Jun-12-13      &   1766.00 &   13.7  &    1.5  &   0.75  \\
2003-Sep-10-11      &   1921.50 &   16.5  &    1.8  &   1.02  \\
2003-Oct-21-22      &   1684.83 &   27.4  &    3.0  &   0.89  \\
2004-Jan-03-05      &   2248.34 &   22.8  &    2.4  &   0.87  \\
2004-Mar-15-17$^b$  &   2279.67 &   25.8  &    2.8  &   0.91  \\
2004-May-11         &   1105.00 &   15.7  &    1.9  &   1.11  \\
2004-Nov-16-17      &    861.00 &    6.3  &    0.8  &   0.93  \\
2004-Nov-27.11      &    519.17 &    5.2  &    0.7  &   1.00  \\
2005-Feb-16-17$^b$  &    781.66 &    7.2  &    0.8  &   0.90  \\
2005-Mar-26$^{b,c}$ &    801.17 &   49.5  &    5.4  &   0.59  \\
2005-Jul-24-25$^c$  &    575.33 &   30.6  &    3.3  &   0.53  \\
\hline
\multicolumn{5}{l}{\it Average values:}\\
all sessions &          &$17.4\pm 2.4$&   &$0.92\pm 0.05$ \\
w/o last two   &          & $16.0\pm1.7$&   & $0.96\pm 0.04$\\         
\hline\hline
\multicolumn{5}{l}{{\bf Notes:} $^a$ both scintillation timescales observed}\\
\multicolumn{5}{l}{\phantom{\bf Notes:} $^b$ structure function does not saturate for this epoch}\\
\multicolumn{5}{l}{\phantom{\bf Notes:} $^c$ untypical flux variations - see Fig.\ref{fig1}d}\\
\hline
\end{tabular}
\end{table}


\subsection{Individual session flux density variations. \label{flux_ind}}

The modulation index $m$ is defined as the ratio of the RMS deviation
to the mean value of the observed flux densities 

\begin{equation}
\label{mod1}
m=\frac{\sqrt{\left< \left(F - \left<F \right>\right)^2 \right>}}{\left<F\right>}~.
\end{equation}

Table~\ref{t1} presents the values of the measured modulation indices for each 
individual session. The average values of $m$ in the bottom section of the table, just as the flux 
density averages (see previous section), are calculated with and without last two sessions. 
For these sessions the quasi-stable flux behaviour means that the relative strength of the modulation 
drops and also causes the increase of $\left<F\right>$. Both these effects contribute to the 
decrease of the measured modulation index for the two ultimate sessions.

For the remaining sessions the modulation index varies from 0.71 (November 2002) to 1.32 (August 2002), 
with an average value of $m=0.96\pm0.04$, which means that the observed flux density of the source is 
undergoing significant variations. One has to ask a question: what contributes to the observed flux modulation?

We expected that at the frequency of 4.8~GHz PSR~B0329+54 will be still in strong scintillation
regime, which means that its signal will be undergoing diffractive as well as refractive scintillations.
This came from the observational experience with this pulsar and strong flux variability observed
at 5~GHz, when we observed it for the purpose of estimating its spectrum (Maron~et~al. \cite{maro00}). The 
theory (see Lorimer \& Kramer, \cite{lori}, for summary) predicts that at 4.8~GHz a pulsar with 
$DM = 26.8$~pc~cm$^{-3}$ should be close to its transition frequency, but on the strong scintillation 
side as well.

For the DISS the intrinsic modulation index should be close to unity. In usual, low-frequency observations, 
where the decorrelation bandwidth $\Delta\nu_{\rm DISS}$ is significantly less than the observing 
bandwidth, the number of observed scintles within the bandwidth, at any given time, will be large. 
That would lead to the decrease of the modulation of the average pulsar flux (integrated over 
the whole bandwidth), as at any given time, one would see multiple scintles, possibly 
in different stages of development.

Our observing bandwidth of 192~MHz is at least four times more narrow than the expected 
$\Delta\nu_{\rm DISS}$ (see Table~3). It means that at any given time we are usually
able to see only a fraction of a single scintle. Following Cordes~\&~Lazio~(\cite{cordes91}), 
we estimated our number of scintles in both frequency and time, and calculations yielded $N_f$
to be very close to unity (as expected) and $N_t \simeq 4.6$. The  latter leads to the expected
DISS modulation index of 0.466, much lower than the observed values. However, after including the 
contribution from RISS ($m_{\rm RISS}=0.56$; the value found via structure function analysis, see 
next sections of the paper), and using the total intensity variance formula (Rickett \cite{rick}),
we obtained the final expected value of $m_{\rm tot}=0.889$. This value is still somewhat lower than the 
observed modulation indices, but not by a huge margin.


\subsection{Average flux density variations \label{flux_av}}

Because every individual session was at least several hours long, we believe that diffractive scintillations
(which happen at the timescale of ca. 40 minutes, see subsection~\ref{sect_sf}) should not affect 
our average flux measurements. On the other hand, refractive timescales, which are significantly longer, may 
have affected our results.

\begin{figure}
\vspace{1.2cm}
   \centering
   \includegraphics[width=8cm]{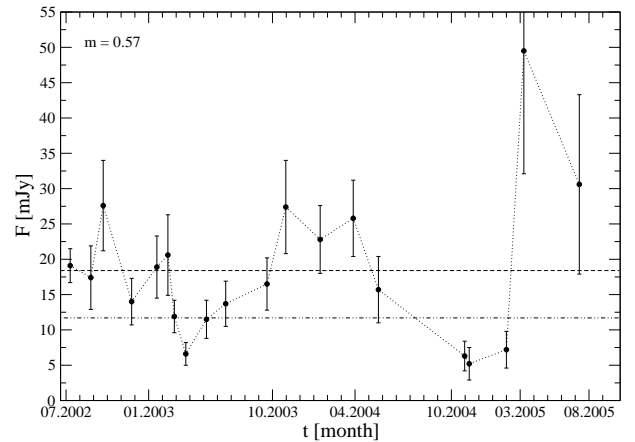}
      \caption{Flux density for each epoch of flux density monitoring
              at 4.8 GHz. The error of the mean flux density was
  calculated from $\sigma F/F\approx m_{\rm RISS}/(t_{obs}/t_{\rm
              RISS})^{0.5}$ (KS92). The top horizontal line represents
             the arithmetic average flux, the lower line corresponds to the weighted average flux.\label{flux_long}}
   \end{figure}

Figure~\ref{flux_long} shows the results of the average flux density measurement versus the observing epoch.
Clearly (conf. Table~1) there is significant variation in the average values. Using these data, we 
calculated the modulation index of the average flux density measurements, which yielded the value of 
$m=0.57$. Assuming that the pulsar itself is not varying in intensity, this modulation could be caused by only 
the refractive scintillations, which happen at significantly long timescales, close to the length of a 
single observing session.

To take that into account, we calculated the errors of average flux measurements 
following Kaspi \& Stinebring (\cite{kasp}, KS92) as

\begin{equation}
\frac{\delta F}{F} \simeq \frac{m_{\rm RISS}}{\sqrt{T_{\rm obs}/t_{\rm RISS}}},
\end{equation}

\noindent
where $T_{\rm obs}$ is the length of a given observing session, and $t_{\rm  RISS}$ is the RISS 
timescale. To calculate the error values shown on Figure~\ref{flux_long} we used the value
of $t_{\rm RISS}$ obtained via the structure function analysis (305 minutes, see 
subsection~\ref{sect_sf}).

Using those error estimates, we were able to calculate the weighted average flux density ($w_i = \sigma_i^{-2}$)
for our entire observing session, which is $\left<F_{\rm tot}\right> = 11.69$ mJy. This value is shown in 
Figure~\ref{flux_long} as a dashed-dotted horizontal line, along with the simple arithmetic average from 
Table~\ref{t1} (dashed line).

As we mentioned above,  the average flux values and their respective uncertainties for the last two 
sessions in our project differ significantly from the remaining observations. However, a proper calculation
of the error estimates improves the picture. For relatively short sessions, which 
we had towards the end of the project, RISS can play a huge role, but the duration of the session is included
in the uncertainty estimates, which makes them more reliable than the formal errors cited in Table~1.
The weighted average is therefore much lower than the arithmetic average.

At this point one has to note that the estimation of the flux density of pulsars at high observational
frequencies will always be affected by strong scintillations (except for some very low DM cases) and 
one has to take it into account when trying to measure the flux of these, for example for the purpose 
of acquiring pulsar spectra.


\subsection{Structure function analysis: timescales\label{sect_sf}}

To estimate the scintillation timescales, we used the structure function analysis of the flux density variations.
For uniformly sampled data the first order of the normalized structure function of the flux density is 
(Simonetti~et~al. \cite{simo})

\begin{equation} 
D(j)=\frac{1}{\left<F \right>^2 N(j)} \sum w(i)w(i+j)[F(i) - F(i+j)]^2,
\end{equation}

\noindent
where $i$ is the data index, $j$ is the offset that corresponds to a time lag in the data for which 
the value of the structure function is calculated. $w(i)$ is the weighting factor (which is equal to 
1 where the flux density exist for $j$th points and 0 otherwise), and $N(j)=\sum w(i)w(i+j)$ is 
the number of pairs of values found in the data that were found to have the offset equal to $j$ 
(i.e. time lag equal to $\tau$).\\

The values of the structure function need to be corrected to remove white noise 
contribution. This was made by subtracting the value of the 
structure function at the unit lag from all the structure function values.
If a quasi-periodic signal is present in the data the structure function will show a plateau at
the value of $D_{\rm sat}$. The timescale of the observed variability is the lag 
corresponding to half the saturation value of the structure function
after white noise correction: $t_{\rm ISS}=\tau(D_{\rm sat}/2)$ (KS92).

The error of the timescale can be estimated by finding the time lags that correspond to
the values of the structure function of $\left(D_{\rm sat}\pm\delta D_{\rm sat}\right)/2$. 
The uncertainty of the saturation value of the structure function
$\delta D_{\rm sat}$ is best estimated as $\delta D_{sat}/D_{sat} \approx (2t_{\rm ISS}/T_{\rm obs})^{1/2}$,
where $T_{\rm obs}$ is the total duration of the observation, in our case the length of an 
individual observing session.

Since our project consisted of several observing sessions repeated over the course of three
years, we decided to calculate the general average structure function as well. 
This can be made either directly by applying structure function algorithm to the entire set of data, or by adding 
the values obtained for individual sessions, after de-normalizing them.

The values of the scintillation timescale, modulation index and their uncertainties for the general average
structure function can then be calculated in the same manner as for the individual sessions.

\begin{table}[t]
\caption{Parameters calculated from structure function for flux density series: 
scintillation timescales and flux modulation indices. The top part of the table presents the values for DISS
timescales and modulation indices for individual sessions, as well as mean values and results from 
general average SF analysis. The bottom part presents RISS parameters for two individual sessions
and the general average SF.\label{sf_table}}
\begin{tabular}{lcrllc}
\hline\hline
Date~~~~~~~~~   &  Lag & $t_{\rm ISS}$~~~ & ~~$\sigma(t_{\rm ISS})$ & ~~$m$ &$\sigma_{m}$  \\       
                & (min)& (min)~          &  ~~(min)                &       &              \\
\hline
2002-Jul-03-08   &  6300 &   58.3         & $^{+5}_{-6}$    & 0.95 & 0.15   \\
2002-Aug-22-23   &  1454 &   45.0         & $^{+12}_{-13}$  & 1.23 & 0.34     \\  
2002-Sep-18-19   &  1816 &   29.0         & $^{+4}_{-4}$    & 0.89 & 0.2     \\
2002-Nov-19-20   &  1731 &   33.4         & $^{+5}_{-4}$    & 0.70 & 0.2   \\ 
2003-Feb-06-07   &  1227 &   45.8         & $^{+11}_{-10}$  & 0.81 & 0.29   \\   
2003-Feb-19-21   &  2650 &   33.7         & $^{+6}_{-5}$    & 0.77 & 0.16   \\  
2003-Mar-16-18   &  1707 &   32.4         & $^{+8}_{-6}$    & 0.82 & 0.21   \\   
2003-Jun-12-13   &   861 &   29.5         & $^{+6}_{-5}$    & 0.80 & 0.27  \\   
2003-Sep-10-11   &  1921 &   58.1         & $^{+17}_{-20}$  & 1.00 & 0.29  \\   
2003-Oct-21-22   &  1684 &   81.6         & $^{+14}_{-20}$  & 0.91 & 0.35  \\ 
2004-Jan-03-05   &  1489 &   34.5         & $^{+3.5}_{-7}$  & 0.86 & 0.23   \\ 
2004-May-11      &  734  &   37.0         & $^{+18}_{-10}$  & 0.90 & 0.36  \\
2004-Nov-16-17   &  861  &   37.4         & $^{+7}_{-5}$    & 0.93 & 0.34  \\   
2004-Nov-27      &  519  &   25.4         & $^{+13}_{-6}$   & 1.02 & 0.38  \\
2005-Jul-24-25   &  575  &   19.0         & $^{+5}_{-4}$    & 0.49 & 0.21    \\
\hline
\hline
mean values  &      &  $40.0$ & $\pm 4.2$ & $0.88$&$\pm 0.05$   \\ 
\hline
average SF  &       &   42.7         & $^{+4}_{-5}$  & 0.89 & $\pm0.06$  \\
\hline\hline
\multicolumn{6}{c}{RISS Parameters} \\
\hline
2002-Jul-03-08 &  6300 &   363        & $^{+7}_{-14}$  & 0.48 & 0.28  \\
2002-Sep-18-19 &  1219 &   326        & $^{+28}_{-22}$ & 0.71 & 0.59   \\
average SF    &       &   305        & $^{+40}_{-15}$ & 0.56 & 0.12 \\
\hline\hline
\end{tabular}
\end{table}

Table~\ref{sf_table} shows the results of the structure function analysis applied to our data. This table
has less individual session entries than Table~1 because it shows only those sessions for 
which the structure function saturated. The saturation of the SF was necessary to calculate the cited values of 
$t_{\rm ISS}$ and $m$ and their uncertainties (see the notes for Table~1).

\begin{figure}

\ \put(0,170)
{\includegraphics{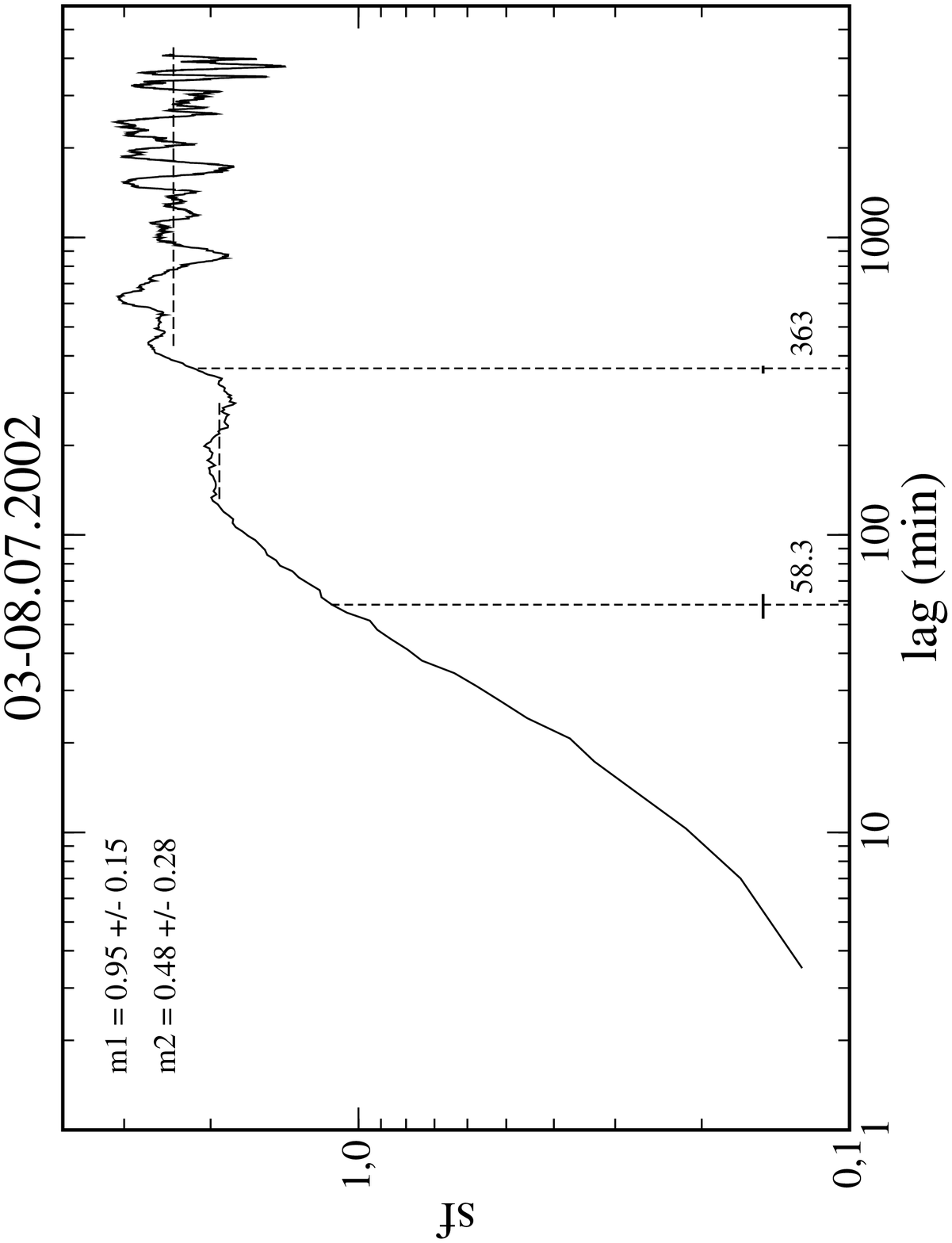}}
\ \put(0,-10)
{\includegraphics{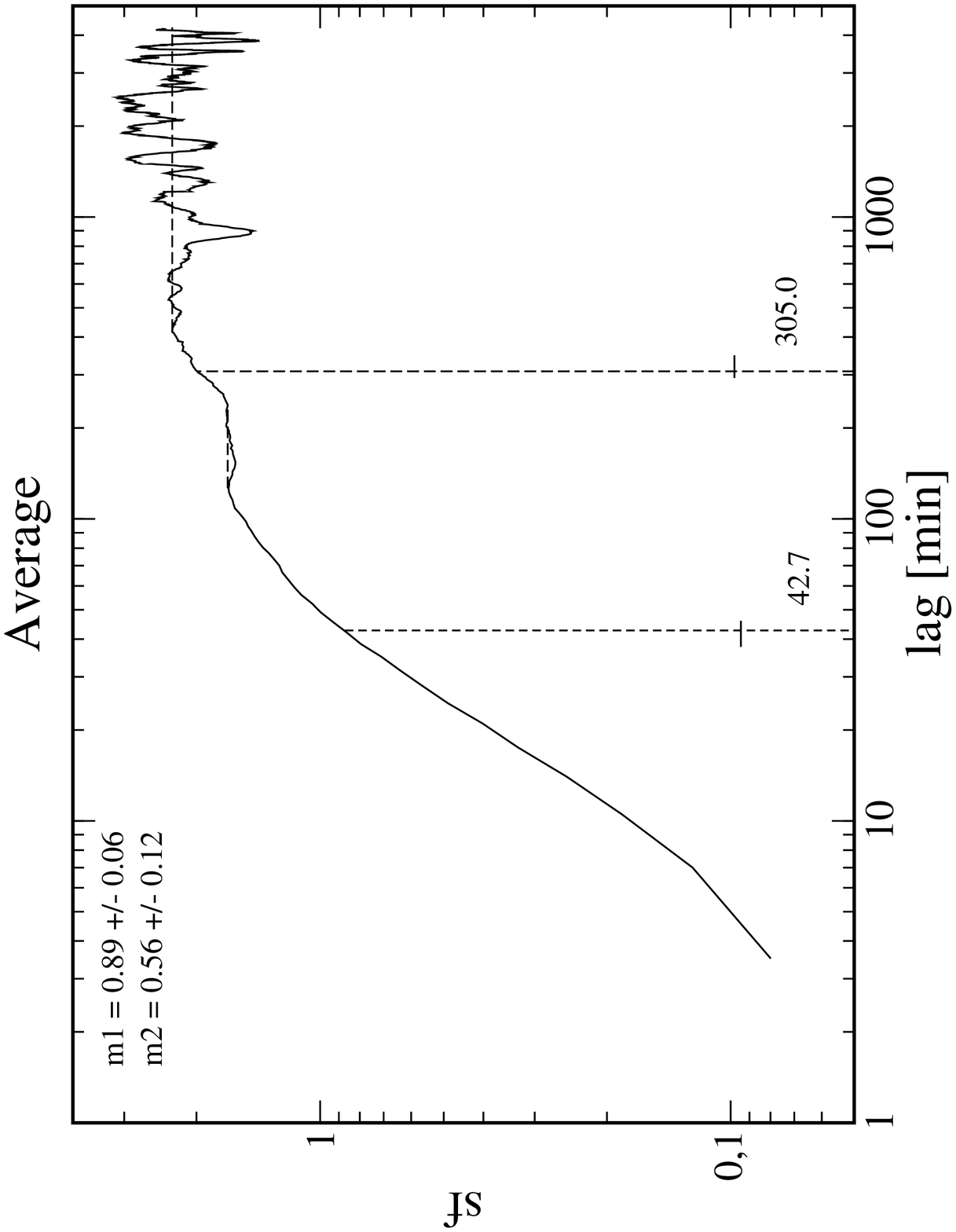}}
\ \put(0,-190)
{\includegraphics{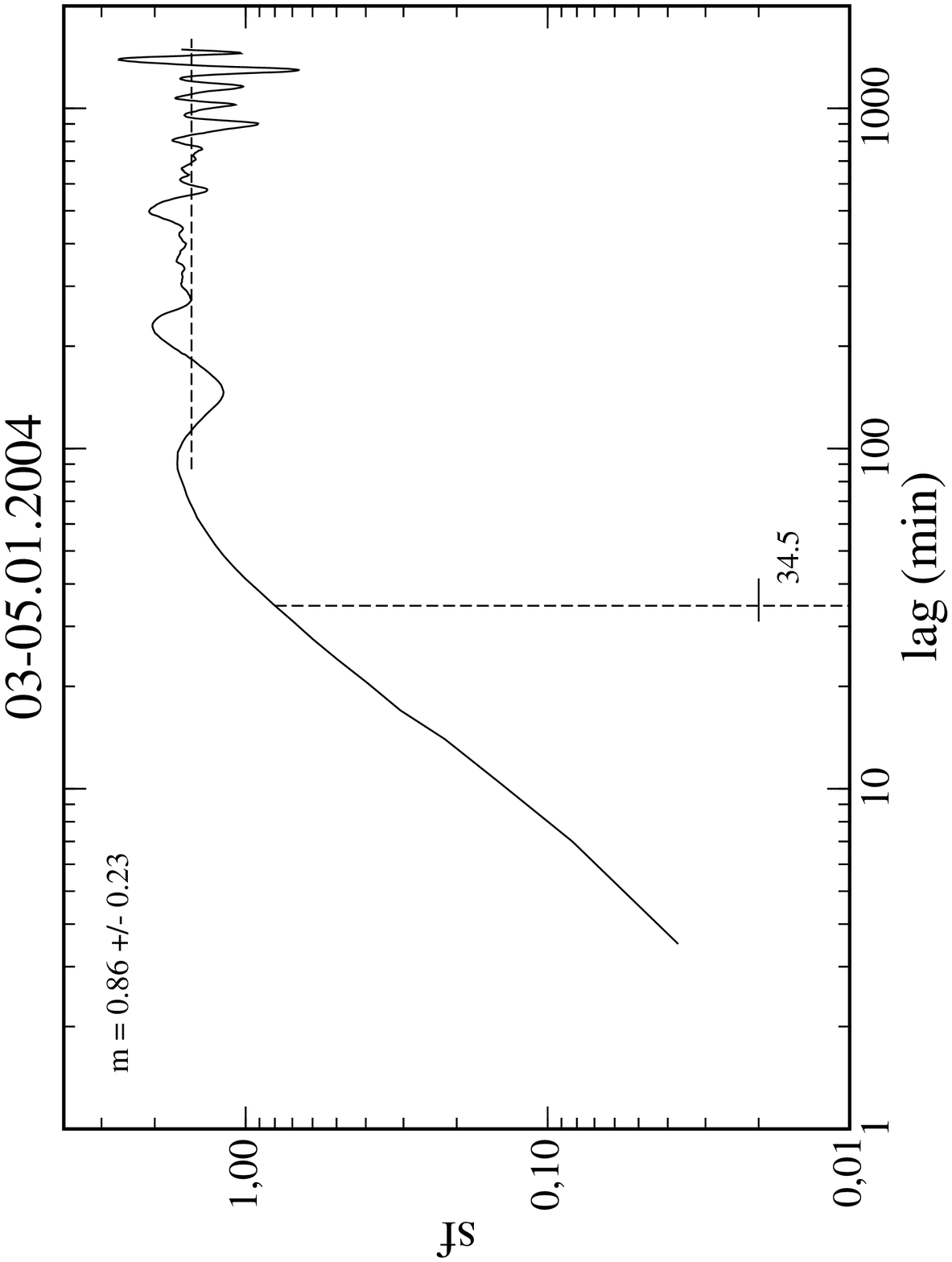}}
\vskip70mm
      \caption{Three structure functions of our data. From top to bottom: the structure function 
              from the 5-day session, general average structure function, and the typical structure 
              function of an individual session (see text for details).   \label{fig2}}
   \end{figure}

Figure~\ref{fig2} shows three of the structure functions obtained during the analysis. The top plot shows the SF
for the first and longest, 5-day session that was conducted between 2002 July 3-8. This structure 
function clearly shows two plateaus, which allowed us to obtain the values of two distinctive timescales present 
in the data. As we mentioned above, at the observing frequency of 4.8~GHz PSR~B0329+54 is believed to be 
close to the transition frequency, i.e. switching from strong to weak scintillation regimes,  
diffractive and refractive timescales should be relatively close (yielding a low value of the strength of 
the scattering parameter $u$). Hence we are convinced that those two plateaus correspond to two 
scintillation timescales, a shorter diffractive timescale (lower plateau), and a longer refractive timescale.

This is even more convincing for the middle plot of Fig.~\ref{fig2}, which shows the general average 
structure function. Two plateaus were also present for another session, 2002 September 18-19 (ca. 30-hour session).

The third and bottom subplot of Fig.~\ref{fig2} shows a typical structure function, obtained during a single, typical 
37-hour session, where one can see only the regular saturation of the structure function caused by the 
diffractive scintillations.

The horizontal lines in all those plots show the SF saturation levels, vertical lines - the time lags that 
correspond to the saturation (the values of the timescales are given in minutes), while the short horizontal 
dashes crossing them represent the errors in the timescale estimates.

All detailed results are presented in Table~\ref{sf_table}. The values of $t_{\rm RISS}$ for 
individual sessions differ from the average SF values. 
This is understandable because the RISS saturation level is determined with lower precision, 
because it is limited by the maximum time lag of the SF calculated for the given session, and because 
especially in the September 2002 we saw only 4-5 RISS cycles. This translates into a large uncertainty 
of $t_{\rm RISS}$. While for the September 2002 session the values agree with those obtained from the general 
average SF within error estimates, for the first and longest July 2002 session they do not. This may be because 
this session may also suffer from the untypical flux variations that were observed (see Figure~1 and 
Section~\ref{flux}).

Because the general average structure function of our data is the best representative for our results, we'll be using
those values for further calculations: $t_{\rm DISS} = 42.7$ minutes, $t_{\rm RISS} = 305$ minutes. Given 
these, we can calculate the strength of the scattering parameter $u= \sqrt{ t_{\rm RISS}/t_{\rm DISS}} = 2.67$. 

Also, since in our observations we were unable to obtain the value of the decorrelation
bandwidth $B_{\rm DISS}$ (this value greatly exceeds the total observing bandwidth of PSPM~2, which is 192~MHz) 
we used the results of our timescale measurements to estimate this value. Using the Stinebring \& Condon (\cite{stin}) 
formula:

\begin{equation}
\label{bw_stine}
B_{\rm DISS} = 1.27 f_{\rm obs} \frac{t_{\rm DISS}}{t_{\rm RISS}},
\end{equation}

\noindent
we obtained a value of $B_{\rm DISS} = 853$~MHz (over 4 times higher than the observing bandwidth).


\subsection{Modulation indices from the structure function analysis\label{modul_sf}}

As we mentioned in the previous section, the structure function analysis allows us to calculate not 
only the scintillation timescales, but also their respective scintillation indices. For a normalized and 
white-noise corrected structure function the value of the scintillation index $m=(D_{\rm sat}/2)^{1/2}$, 
and the fractional error in $m$ is a sum of errors from the fractional errors in $D_{\rm sat}$ and $m$ (see KS92).

Table~\ref{sf_table} shows the results for the individual sessions as well as for the general average SF for
the modulation indices of both diffractive and refractive scintillations.

For the diffractive scintillations the average value of $m_{\rm DISS}$, obtained from the simple
arithmetic average of the individual values agrees with the value obtained from general average structure 
function, which is equal to $m_{\rm DISS}=0.89$. This value differs slightly from the modulation index value 
obtained directly from flux density measurements (0.96, see Table~1), but is very close to the expected value
for the observed modulation index (0.889, see section \ref{flux_ind}).

The values of the  RISS modulation indices are also presented in Table~\ref{sf_table}. For individual sessions
that allowed for an estimation of $m_{\rm DISS}$, they differ slightly from the average values, but still agree 
within the error estimates. This is not a surprise, because it is 
difficult to determine the saturation level properly, especially for the shorter sessions (for the same 
reasons the timescales differ, see previous section). On the other hand, the value of  $m_{\rm RISS}=0.56$ 
obtained for the general average structure function is very close to the value of the RISS modulation
index obtained from the average flux density variation analysis (0.57; see section~\ref{flux_av} and 
Fig.~\ref{flux_long}). 

Again, as in the case of the ISS timescales, we will be using the values obtained for the general average structure 
function as the final result of our analysis:  $m_{\rm DISS}=0.89$ and  $m_{\rm RISS}=0.56$.


\section{Discussion}

The results obtained from the analysis of our observations suggest that we definitely see both refractive 
and diffractive scintillations affecting the measured pulsar flux density. We were able to calculate both the 
timescales and modulation indices for these, and also to estimate the expected decorrelation bandwidth. 

Our observations were performed at the frequency of 4.8~GHz, for which there is only one observational 
report to be found in the literature by Malofeev at al.~(\cite{malo}). This was quite a short project, because
PSR~B0329+54 was only observed for 130 minutes at this frequency, which given the values of the scintillation 
timescales we found makes the results unreliable. On the other hand, adding our results to the scintillation 
parameters estimates obtained by other authors at lower frequencies allows us 
to study the frequency dependence of these parameters over a much wider range of frequencies.

\subsection{Flux density measurements}

As we mentioned above, PSR~B0329+54 at the observing frequency of 4.8~GHz shows large flux 
density variations owing to ISS. For some of the individual 3-minute integrations the flux measurements yielded 
values as high as 200~mJy, over 10 times higher than the average value. Those variations 
happen at a variety of timescales, 
ranging from minutes to hours, and up to days (c.f. Fig~1a). This clearly shows the dangers of estimating pulsar flux 
densities at high frequencies (especially for low-DM pulsars) from observations that are based on a limited number
of separate integrations, which is often the case when one tries, for example, to ascertain the pulsar spectrum.
The diffractive timescale is longer at high frequency, and may be comparable to the actual integration times used
- usually of the order of a few tens of minutes when one tries to measure the flux density of some weaker 
sources - hence, it may strongly affect the outcome. On the other hand, 
with refractive timescales of the order of a few hours, 
repeating the observations after a day or a few days, with the hope that the ISS will be averaged-out, 
may not solve the problem. One has to approach these cases carefully; luckily, for high-DM pulsars 
this should be less of a problem, because the transition frequency will be much higher 
and the modulation caused by RISS should be very low, and 
one has to worry only about diffractive scintillations when performing flux measurements.

In case of our observations of PSR~B0329+54 we were able to ascertain the impact of interstellar scintillations on
our measurement pretty well. Both our values: 18.4~mJy from the simple arithmetic average and 11.7~mJy from the RISS 
weighted average agree well with the simple power-law spectra for this source, with spectral index of 
-2.2 (see Maron~et~al. \cite{maro00}, and the authors on-line spectra database at 
{http://astro.ia.uz.zgora.pl/olaf/paper1/index.html}).

\subsection{Scintillation parameters \label{scint_params}}

\begin{figure*}
\ \put(0,170)
{\includegraphics{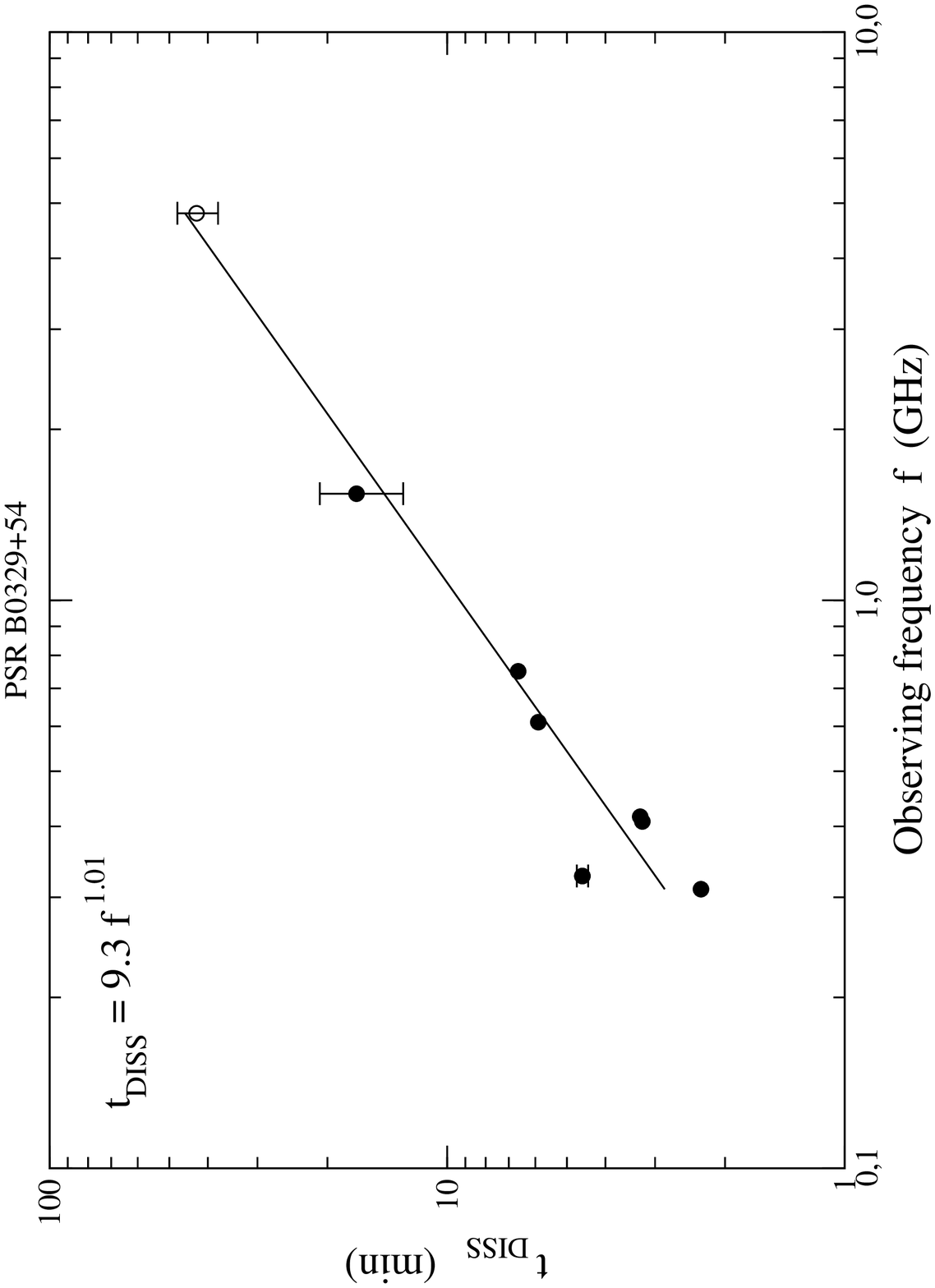}}
\ \put(220,170)
{\includegraphics{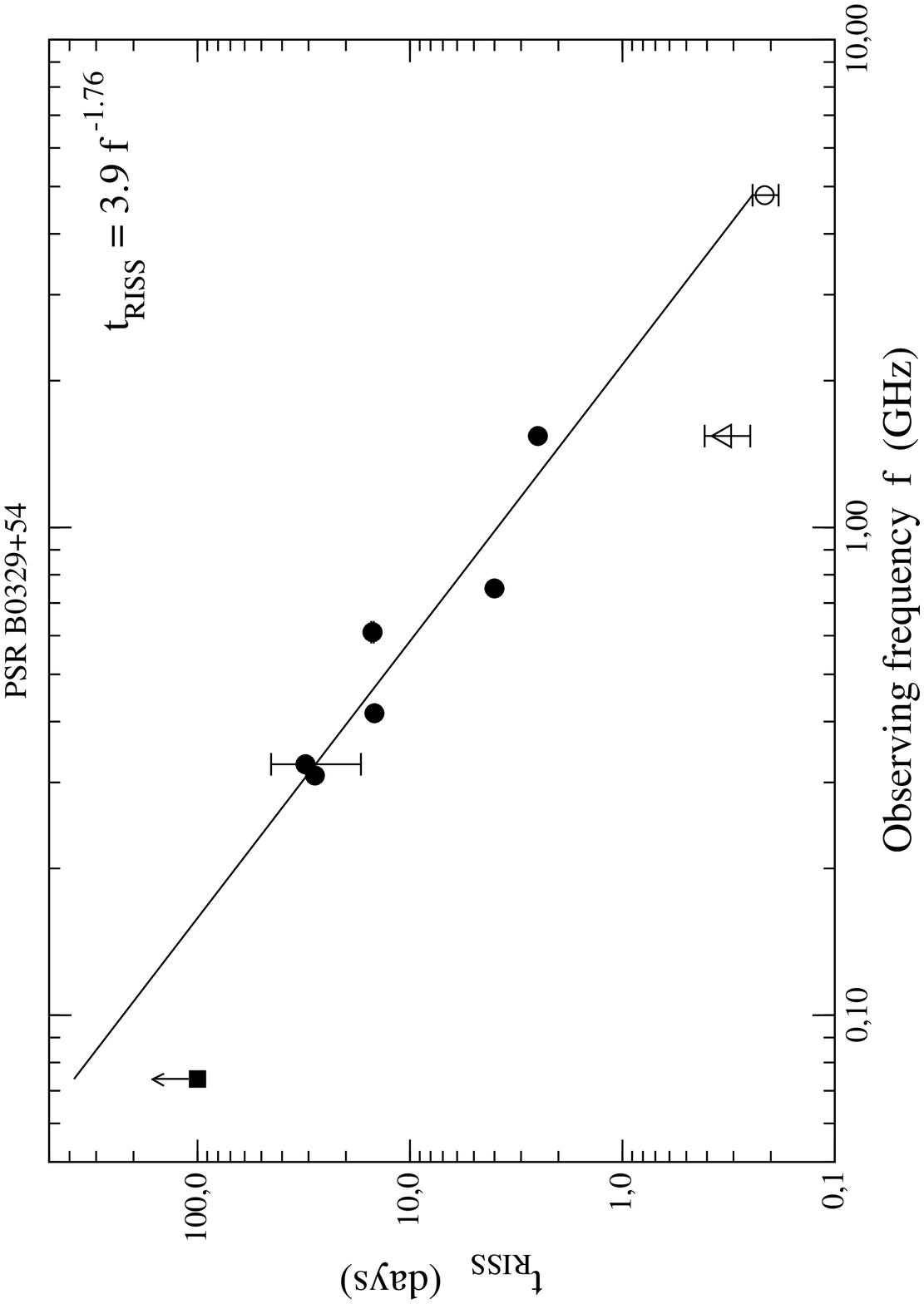}}
\ \put(0,-5)
{\includegraphics{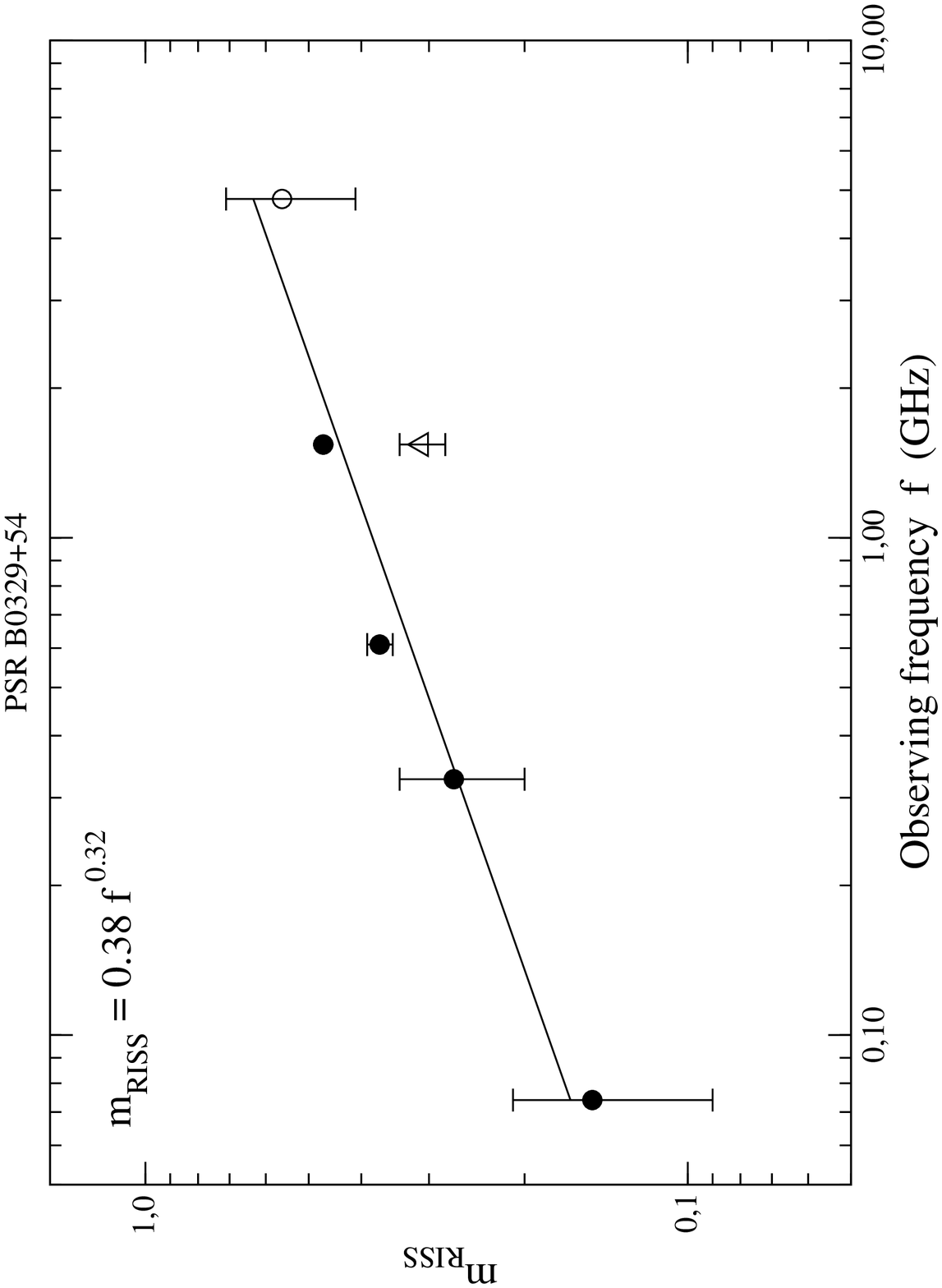}}
\ \put(220,-5)
{\includegraphics{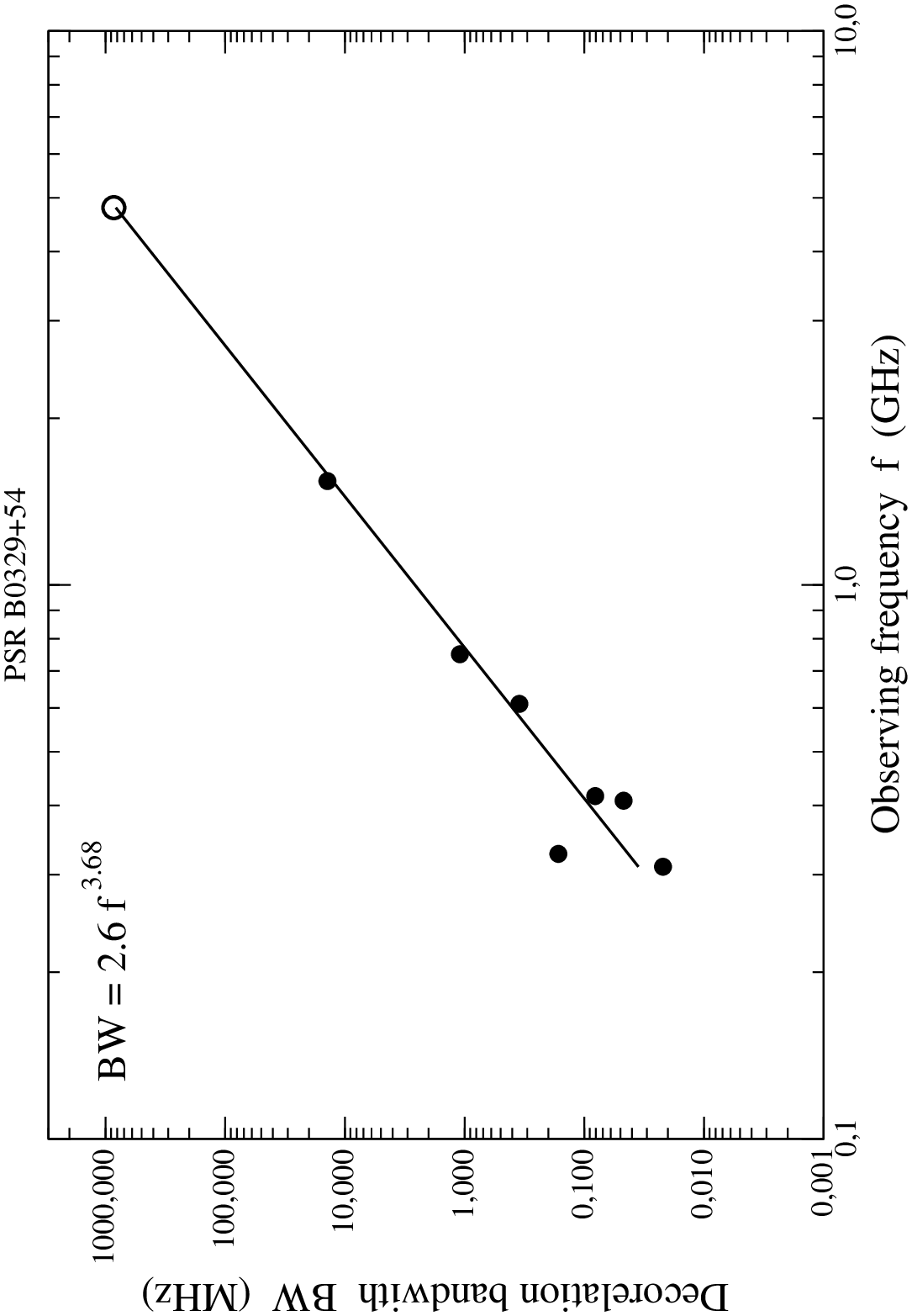}}
\vskip63mm
\caption{Frequency dependence of diffractive timescales $t_{\rm DISS}$ (top left),
refractive timescales $t_{\rm RISS}$ (top right), refractive modulation index $m_{\rm RISS}$ 
(bottom left), and decorrelation bandwidth $BW$ $m_{\rm RISS}$ (bottom right).
Data taken from Bhat~et~al. (\cite{bhat}) at 327 MHz, Gupta~et~al. (\cite{gup94}) at
408 MHz, Stinebring~et~al. (\cite{stin96}) at 610 MHz, Wang~et~al. (\cite{wang}) at
1.5 GHz, and this paper at 4.8 GHz.
The solid line presents the fit to observed data. \label{fig4}}
\end{figure*}

Pulsar B0329+54 is one of the strongest radio pulsars known so far, and scintillations of its radiation were 
analysed at various frequencies by many different authors, especially over the course of last 20 years. 
Table~\ref{tab3} and Figure~\ref{fig4} summarize the results we were able to find.

Table~\ref{tab3} shows the list of all scintillation parameters for PSR~B0329+54 at different frequencies 
ranging from 74~MHz to 4.8~GHz. At 4.8~GHz, at which we conducted the observations
as well, we only found one measurement of $t_{\rm DISS}$ in Malofeev~et~al.~(\cite{malo}), which is 21.4 minutes, 
almost exactly two 
times lower than our value from the general average structure function. This discrepancy may arrise because
 Malofeev~et~al.~(\cite{malo}) observations were very short: 130 minutes only, i.e. three diffractive cycles, according 
to our results. Additionally, one has to note that even during our observations we found that for  individual 
sessions the measured value of $t_{\rm DISS}$ can go as low as 19.0 minutes  (see Table~\ref{sf_table}).
Malofeev~et~al.~(\cite{malo}) also estimated a lower limit for the diffractive modulation index ($m_{\rm DISS} > 0.5$).

\begin{table*}
\caption{Estimated and observed parameters. Numbers in parentheses are error estimates 
at the level of the last digit(s) quoted. \label{tab3} }
{
\begin{tabular}{lcccccccl}
\hline \hline
                      & $t_{\rm DISS}$ & $t_{\rm RISS}$ & $m_{\rm DISS}$ & $m_{\rm RISS}$ & $u$ & $B_{\rm DISS} $ & $v_{\rm ISS}$ & References   \\
                      &  (min)     & (min) & & &    & (MHz) & (km/s) & and notes \\ \hline \hline
{\bf Our 4.8~GHz Results}         & 42.7 & 305 & 0.87 & 0.56 & &&&\\
derived parameters    &      &     &      &      & 2.67 & 853 & 92.2&  \\ 
\hline
\multicolumn{9}{l}{\bf Results from literature:}\\ \hline
4.8~GHz  & $21(4)$ & & $>0.5$ & & & & & 1 (short observations) \\ 
1540~MHz    &  & (?)480 &  & (?)0.31 & & & & 2 (doubtful)\\
1540~MHz     & 16.9  & 3600 & 0.47 & & 15 & 14.0 & 97 & 3\\ 
750~MHz & 6.64 & 4.0 days & & & & 1.096 & &4 \\
610~MHz  & 5.9 & 15 days & & $0.37(3)$ & & 0.349&  & 5,6\\ 
416~MHz & 3.27 & 14.7 days & & & & 0.081& & 4 \\ 
408~MHz & 3.23 & & & & & 0.047& & 7\\ 
327~MHz  & 4.57 & & & & 63 & $0.165(8)$ & & 8\\ 
327~MHz  &  & $31(14)$ days & & $0.27(7)$ & & & &9 \\ 
310~MHz  & 2.3 & 28 days & & & & 0.0022 & & 4\\ 
74~MHz & & $>100$ days & & $0.15(6)$ &&& & 7 \\ \hline \hline
\end{tabular}
\vskip1mm
{{\bf References:} (1)~Malofeev~et~al. (\cite{malo}), (2)~Wang~et~al. (\cite{wang08}), (3)~1540~MHz Wang~et~al. (\cite{wang}),
(4)~Stibenring \& Condon (\cite{stin}), (5)~Stinebring~et~al. (\cite{stin96}), (6)~Stinebring~et~al. (\cite{stin00}),
(7)~Gupta~et~al. (\cite{gup94}), (8)~Bhat (\cite{bhat}), (9)~Esamdin~et~al. (\cite{esam}) }
} 
\end{table*}

Figure~\ref{fig4} shows the frequency dependence for both diffractive and refractive scintillation timescales
(top right and top left, respectively). Data taken from the literature are represented by full circles (for 
reference see Table~\ref{tab3}), our measurements are the open circles. Evidently, our measurements for both 
$t_{\rm DISS}$ and $t_{\rm RISS}$ roughly agree with the previous, lower frequency estimates
in a sense that they support the theory that the frequency dependence of the scintillation timescales is really
governed by a simple power-law. Adding our points to these plots extends the frequency range significantly, which
in turn allows for better fit of the power-law to the data. This is especially important for the RISS timescale, 
because it is very long at lower frequencies and the measurements 
often suffer because they are  either poorly sampled or 
have only short time spans; i.e. do not cover a sufficiently large number of observed refractive cycles. 
These result in a large spread of $t_{\rm RISS}$ measurements at lower frequencies.

On the other hand, our measurements were obtained from the continuous observations that usually covered at least 
4-5 refractive cycles (in a normal one-day session) and up to over 20 cycles (for the longest 5-day session), 
which makes them very reliable. 
 
Using the data summarized in Table~\ref{tab3} we performed power-law fits, which yielded 
$t_{\rm DISS} = 9.3 \ f_{\rm obs}^{~1.01}$~(minutes) and 
$t_{\rm RISS} = 3.9 \ f_{\rm obs}^{~-1.76}$~(days).

Similar fits were performed for the RISS modulation index (see bottom left plot in Fig.~\ref{fig4}) and the 
decorrelation bandwidth $B_{\mbox{\tiny DISS}}$, although the latter is not a direct measurement but only 
an estimate (see Section~\ref{sect_sf}). Nevertheless, both measurements roughly agree with expectations based
on lower frequency data, and adding our points to the pool allowed us to obtain the frequency dependencies as
$m_{\rm RISS} = 0.38 \ f_{\rm obs}^{~0.32}$ and 
$B_{\rm DISS} = 2.6 \ f_{\rm obs}^{~3.68}$.

We have to note that for the purpose of two of these fits, namely $t_{\rm RISS}$ and $m_{\rm RISS}$,
we decided to omit the values obtained by Wang~et~al.~(\cite{wang08}), which are represented by triangles in their respective 
plots. These points stray by much from the rest of the data, and contradict the earlier estimates at the same 
frequency (1540~MHz) given by Wang~et~al. (\cite{wang}). Adding our values to the data pool 
clearly shows that their earlier results (obtained from a different set of observations)
seem to fit better into the general picture. We believe that the 2008 results are flawed; 
problem may lie in the method
that was used for the purpose of that paper. Based on other data, the actual refractive 
timescale at 1540~MHz should be of the order of 2 days (similar value as that given by 
their earlier paper, which is 2.5 days). Their observations were 
conducted almost continuously for 20 days in March 2004, with the flux measurements based on 90-minute integrations. 
At the frequency of 1540~MHz this means that they observed only a few diffractive cycles during their single 
flux-integration, which may have affected the measurements. 
The authors calculated $m_{\rm RISS}$ based on that data. However, their structure function for the flux measurements
did not saturate. Without the saturation level it is impossible to estimate the timescale from the structure function
(see section~\ref{sect_sf}), and they decided to obtain the saturation level as $D_{\rm sat} = m^2/2$ (see 
section \ref{modul_sf}). Using the value obtained this way, and applying it to the structure function 
(see their Fig.~5), they inferred the value of the RISS timescale of 8 hours, 
6 times lower than expected and almost 8 times lower than in their previous paper. 

Probably the flaw lies in the underestimation of the RISS modulation index 
(which clearly shows in our Fig.~\ref{fig4} as well) which resulted in a lowering of the used saturation level. 
An inspection of the structure function they obtained (see their Fig~5, Wang~et~al.~\cite{wang08}) 
leads to the conclusion that even a slight change of $m_{\rm RISS}$, and in turn the saturation level 
$D_{\rm sat}$, could lead to timescale even a half an order of magnitude higher than the 8 hours they quote, 
and we think that is exactly what happened. In Wang~et~al.~(\cite{wang08}) defense we may note that 
we had one session in our data that is coincident with their March 2004 observations, as we performed a 
38-hour observation between March 15 and 
17 of 2004, and our structure function did not saturate as well, which may suggest that the fluctuations
of the flux density were untypical during that period.

Knowledge of frequency dependence of both $t_{\rm DISS}$ and $t_{\rm RISS}$ gives us an opportunity 
to calculate the {\it transition frequency}, i.e. the observing frequency at which both timescales 
would be equal, which
means a switch from strong to weak scintillation regimes. Using the empirical formulae provided above (see also 
Fig~\ref{fig4}) and putting both timescales equal, this yields the transition frequency $\nu_c$ of 10.1~GHz. Below 
this frequency PSR~B0329+54 will be in the strong scintillation regime, above that frequency, the characteristic 
of the scintillation should change to weak scintillations.

\subsection{Estimation of electron density turbulence spectrum}

Observations of the interstellar scintillations and especially the frequency dependence of the measured parameters
can be used to estimate the properties of the interstellar turbulence. By analogy to the neutral gas theory, the
 density fluctuations in the ISM can be described by a power-law spectrum as
$P_{\rm 3n}=C^{2}_{n}~q^{-\beta}$,
where $C^2_n$ is the mean turbulence of electron density along the line of sight, $q=2\pi/s$ is the 
wavenumber associated  with the spatial scale of turbulence $s$, and the spectral index $\beta$, which is 
in range of $3<\beta<5$. The well known Kolmogorov theory developed for neutral gas turbulence yields 
the spectral index of $\beta = 11/3$, but it was shown by several authors (see for example Gupta~et~al. \cite{gupt}) that there are discrepancies between Kolmogorov theoretical predictions and actual 
measurements of scintillation parameters for some pulsars. Few models of the interstellar turbulence
with different spectral slopes $\beta$ were developed, yielding different sets of predictions of the 
frequency dependence of ISS parameters (see Romani~et~al. \cite{romani}; Bhat~et~al. 
\cite{bhat}). Table~\ref{indx_table} summarizes these predictions for the three most commonly used spectral slopes of
$\beta = $ 11/3 (Kolmogorov spectrum), 4 and 4.3 for the four scintillation parameters we were able to measure, 
or derive from our observations of PSR~B0329+54. Our estimations of the actual spectral slopes of these
parameters, based on all available data (see section \ref{scint_params} and Figure~\ref{fig4}) are also in the table.

Apparently none of the currently models for the electron turbulence agrees precisely with our estimations, 
and the model predictions with $\beta=4$ are the closest to the slopes we obtained from our fits. It has to 
be noted that in case of three of the four parameters (with the $m_{\rm RISS}$ being the only exception), these 
discrepancies cannot be attributed to poor quality of frequency slopes fits for these parameters, especially that 
our measurements, made at the frequency of 4.8~GHz, significantly widened the frequency range, putting strong 
constraints on the fits.

\subsection{Derived scintillation parameters}

As we mentioned above, in a strong scintillation regime the flux density variations happen at two distinctive
timescales - diffractive and refractive. The apparent variability of the pulsar signal can be understood as the 
result of the observer crossing the diffraction pattern, which is introduced by the ISM to the pulsar signal wavefront.
In a thin screen model, those two timescales correspond to two spatial scales of the diffraction pattern: 
the diffractive scale $s_d$, and refractive scale $s_r$, which in turn can be bound to the concepts of 
scattering angle and scattering disk that is commonly used in the scintillation theory (see Rickkett, \cite{rick}).

\begin{table}
\caption{Theoretical and observed spectral indices of scintillation parameters (Romani~et~al. \cite{romani}; Bhat~et~al.~\cite{bhat})\label{indx_table}}
\begin{tabular}{l c c c c}
\hline
 & \multicolumn{3}{c}{\phantom{XXX}Spectral index predicted\phantom{XXX}} & Observed \\
Parameter &  \multicolumn{3}{c}{by theory with $\beta =$} & spectral index \\ \cline{2-4}
  &  \phantom{X} 11/3\phantom{X} &\phantom{X} 4\phantom{X} &\phantom{X} 4.3\phantom{X} & (our work)  \\ \hline \hline
$t_{\rm DISS}$ & \phantom{$-$}1.2  & \phantom{$-$}1.0 &\phantom{$-$}1.4 & \phantom{$-$}1.01 \\
$t_{\rm RISS}$ & $-$2.2 & $-$2.0 & $-$2.4 & $-$1.76 \\
$m_{\rm RISS}$ & 0.57   & 0.38   & 0.55   & \phantom{$-$}0.32 \\
$B_{\rm DISS}$ & \phantom{$-$}4.4 & \phantom{$-$}4.0 & \phantom{$-$}4.7 & \phantom{$-$}3.68 \\ \hline \hline
\end{tabular}
\end{table}

Following Gupta~et~al.~(\cite{gup94}), we can find the diffractive scale by the means of estimating 
the scintillation velocity
$V_{\rm ISS}$, which is bound to the diffractive scale as $s_d=V_{ISS}  t_{\rm DISS}$. The scintillation 
velocity can be derived from the diffractive timescale $t_{\rm DISS}$ and the decorrelation bandwidth
$B_{\rm DISS}$ (Cordes, \cite{cord86}; Gupta~et~al.~\cite{gup94}). Using our values in the formula yields
$V_{\rm ISS} = 92.9^{+7.7}_{-10.4}$ km/s.

The observed scintillation velocity should be modulated by the Earth's orbital motion, but our data 
proved to be inconclusive in that regard. We only had the diffractive timescale measured for 
each epoch, and were forced to use the average value of decorrelation bandwith (and/or refractive timescale)
in the $V_{\rm ISS}$ formula, which led to a wide spread in the $V_{\rm ISS}$ versus epoch data, making 
our fits unreliable. Measurement of the refractive timescale for every epoch would probably 
solve the problem, as would the direct measurement of the decorrelation bandwidth, but these were impossible owing
to the character of the variability of the pulsar flux and the limited observing bandwidth of the backend used.

The average value of $V_{\rm ISS} = 92.9$~km/s combined with the measured $t_{\rm DISS} = 42.7$~min. 
yielded the diffractive scale $s_d = 2.38 \times 10^8$~meters. 

The diffractive angle $\theta_d$ can be estimated from (Wang~et~al.~\cite{wang})

\begin{equation}
\theta_d = \left(\frac{c}{\pi \ d \ B_{\rm DISS}}\right)^{1/2},
\end{equation}

\noindent 
where $c$ is the speed of light. This yields $\theta_d = 0.01207$~milliarcseconds, which translates into 
the refractive scale $s_r = 1.91 \times 10^9$ meters. Combining this with the previously 
obtained value of diffractive scale $s_d$,  we can estimate the Fresnel scale for the observed ISS 
as  $r_F = (s_d s_r)^{1/2} = 6.7 \times 10^8$ meters.

One has to remember that for observations we were not able to measure the decorrelation bandwidth
$B_{\rm DISS}$ directly, and both of the formulae for $V_{\rm ISS}$ and $\theta_d$ are based on this value. 

The value of the strength of the scattering parameter inferred from the spatial scales $s_r$ and $s_d$ is 
$u = 2.84$, which is slightly different from the value derived from the scintillation 
timescales ($u=2.67$, see Section~\ref{sect_sf}). The discrepancy is small, however, and given the fact that the 
application of the ISS theory to the observations always involves lots of approximations and assumptions, 
our values of $s_r, s_d$ and $r_F$ we got are as reliable as possible.

\section{Summary and Conclusions \label{conc}}

We presented the results of our three-year observing project of PSR~B0329+54 at the frequency of 4.8~GHz, using
the 32-meter Toru\'n Centre for Astronomy radiotelescope and the Penn State Pulsar Machine~II as a backend. 
The project was very unique, because it
consisted of twenty separate observing sessions, each of them involving several hours (up to 5 days; more than 
24 hours on average) of continuous observations of the flux density of the pulsar, with 3-minute time resolution.
The total observing time ammounted to {\it 35 thousand minutes}, or {\it 24 days}, making it probably one 
of the longest observing programmes that involved long-term continuous observations of 
flux density and scintillation parameters
that was performed using a single observing setup and uniformly analysed from raw data to the final results. 

\subsection{Scintillation parameters at 4.8 GHz}

The analysis of the data allowed us to conclude that flux density variations are governed by two distinctive
timescales, which we identify as the diffractive and  refractive scintillation timescales. The fact that the
observing frequency of 4.8 GHz is relatively close to the transition frequency for this pulsar makes those timescales
differ only by a factor of $\sim 7$. The character of our observations allowed us to detect them both at the same time,
by means of the structure function analysis, using the same set of data - twice for individual sessions and 
additionally by constructing general average structure function. To our knowledge this is the first case of
pulsar ISS observations where both scintillation timescales were measured from a single data set using exactly 
the same method. 

The measured values of the diffractive timescale $t_{\rm DISS}=42.7^{+4}_{-5}$~minutes and refractive timescale
$t_{\rm RISS}=305^{+40}_{-15}$~minutes allowed us to estimate the decorrelation bandwidth (unobtainable via direct 
measurements, because it exceeds the observing bandwidth of the backend used) as $B_{\rm DISS}= 853$~MHz. Also, 
we were able to estimate the strength of scattering parameter, which can be expressed as square-root of the 
timescale ratio, as $u=2.67$.

We were also able to estimate the amount of modulation that both types of scintillation contribute to the 
flux variability, and did that by the means of the modulation indices. Especially important is the estimation 
of the modulation due to refractive scintillations, for which the modulation index is relatively strong: 
$m_{\rm RISS}=0.56$. This agrees with the fact that during some of our observing sessions we noticed very 
drastic flux density variations, with the flux from single 3-minute integration reaching the values of 200~mJy, 
exceeding the average value by a factor of $\sim 20$.

This shows how important it is to take ISS into account when measuring pulsar flux densities at high frequencies,
especially for low-DM pulsars. For these objects the frequency of the switch from strong to weak
scintillation regimes (i.e. the transition frequency) should be relatively low, and all 
theories predict the increase of refractive modulation index towards that frequency 
(for summary see Lorimer \& Kramer \cite{lori}; also Romani et~al. \cite{romani}).
In these cases, when the ISS timescales are of the order of tens to a few hundred minutes, 
typical flux density observations
(which usually involve continuous integration of the length of 10 minutes to an hour) of a single measurement may be 
heavily affected by both types of scintillations. This in turn may cause problems when trying to construct the 
pulsar spectra, leading to artificial breaks or turn-ups observed in the spectrum - as was pointed 
out in a few cases by Maron~et~al.~(\cite{maro00}). The only way around 
the problem is to repeat the measurement several times to average-out  the influence of ISS. Luckily, this should 
be less of a problem for pulsars with higher dispersion measures because the transition frequency for 
such objects should be
well outside the usual observing frequencies, meaning that at least the refractive scintillations should not 
affect the measurement by much. Still, the diffractive scintillations would present a problem if their timescale 
would be similar to the integration lengths used, so one should approach every case carefully and individually.

Our measurements of the scintillation timescales allowed us also to estimate some of the parameters commonly used 
to describe the theory of interstellar scintillation. We were able to estimate the Fresnell scale $r_F$ as
$6.7 \times 10^8$~meter, which at the frequency of 4.8~GHz corresponds to a refractive scale of 
$s_r = 1.9 \times 10^9$~m, and the diffractive timescale of $s_d=2.4\times 10^8$~m. 

We were also able to estimate the scintillation velocity of $V_{\rm ISS} = 92.9$~km/s, and we hope that 
our value will help to solve the problem with the discrepancies that can be found in the literature. 
It agrees well with Wang~et~al.~(\cite{wang}) as they obtained 97~km/s, on the other hand 
KS92 report $V_{\rm ISS} = 68$~km/s, whereas Gupta~(\cite{gupta95}) 170~km/s or 126~km/s, 
depending on the model. One has to remember that those discrepancies may arise because while 
the the formula is uniform (with the possible exception of the scaling factor $A_v$), the ways used to 
obtain the parameters were usually different.

\subsection{General picture}

Adding our results to the estimations of the scintillation parameters at lower frequencies that can be found 
in the literature allowed us to study the dependence of these parameters on the observing frequency.  
That our observations were conducted at the frequency of 4.8~GHz proved to be very useful because it significantly 
widened the frequency range for which the values of ISS parameters are known. Previous attempts at this analysis
were limited to the frequency of $\sim 1.5$~GHz, adding our measurement expanded that range three-fold, which means
an extension of ca. 50\% when talking about the log-scale in frequency. This proved to be especially helpfull
for the RISS parameters. At low frequencies the RISS timescale is very long and the modulation very weak, 
which brings a relatively large scatter to the low-frequency parameter estimates, making
any attempts to fit a power-law type dependency questionable. The addition of our results, which we believe to 
be very reliable (because they come from a long-term monitoring programme), improves the general picture greatly (see 
Fig~\ref{fig4}), making any fits to the data much more constrained.

We performed the fits of the power-laws governing the frequency dependence of the scintillation parameters, and 
they yielded us the respective spectral indices, which can be expressed as: 
$t_{\rm DISS} \sim f^{~1.01}$,  $t_{\rm RISS} \sim f^{~-1.76}$,
$m_{\rm  RISS} \sim f^{~0.32}$ and $B_{\rm DISS} \sim f^{~3.68}$.
These values significantly differ from those predicted by using Kolmogorov neutral gas theory to the 
ISM spatial electron density spectrum ($\beta = 11/3$, see Table~\ref{indx_table}). Indeed our measurements 
do not agree with any of the models describing the electron density spectrum that are commonly used, with the model
using $\beta = 4$ being the closest to the obtained spectral indices. This would indicate that the real
spatial electron density spectrum is steeper than the Kolmogorov spectrum. Unfortunately, the character 
of our data did not allow for the estimation of the $\beta$ index by the means used by a few other authors
(see for example Wang~et~al.~\cite{wang}) because they usually involve measurements performed in the dynamic spectra.

Using the above mentioned formulae we were able to estimate the transition frequency for PSR~B0329+54, i.e. the 
frequency at which the transition from strong to weak scintillation mode should appear, which is $\nu_c = 10.1$~GHz. 
This clearly contradicts the earlier statement of Malofeev~et~al.~(\cite{malo}), which estimated the transition 
frequency to be around 3~GHz. This value is also higher than the transition frequency one can infer 
from Lorimer \& Kramer (\cite{lori}) - namely from their Figure 4.3, which suggests that for a pulsar with 
$DM=27$~pc/cm$^3$ $\nu_c$ should be close to 4~GHz.

It would be extremely interesting, but on the other hand very hard, to perform an observing project similar to 
ours at a frequency close to this transition frequency. First - the pulsar would be much weaker, thus requiring
longer integration times. On the strong-scintillation-regime side of the transition frequency, the two 
ISS timescales would be extremely similar (100 minutes), which means that the distinction between them
would be very hard to make, and would probably require very long sets of continuous observations. 
Because there are only a handful of telescopes that could perform these observations (i.e. have the 
required receiver equipment) this may be impossible to accomplish.

Nevertheless, any type of scintillation or flux density variation observations, performed at the frequency close
to the transition frequency, may be very helpful to our understanding of the scintillation phenomenon in general, 
and the switch from strong- to weak-scintillation in particular. This is also true for other pulsars, especially
those with lower dispersion measures, because the expected transition frequencies for them would be 
even lower than the 10~GHz for PSR~B0329+54.

\begin{acknowledgements}
We are grateful to G.~Hrynek for his help with  our observations
and  the calibration
procedure. JK, KK and WL acknowledge the support of the Polish State
Committe for scientific research under Grant N~N203~391934.
\end{acknowledgements}

\end{document}